# The collision frequency of electron-neutral-particle in the weakly ionized plasma with the power-law velocity distribution

Futao Sun | Jiulin Du

*Department of Physics, School of Science, Tianjin University, 300072, China*

**ABSTRACT -** We study the collision frequency of electron-neutral-particle in the weakly ionized plasma with the power-law velocity $q$-distribution and derive the formulation of the average collision frequency. We find that the average collision frequency in the $q$-distributed plasma also depends strongly on the $q$-parameter and thus is generally different from that in the Maxwell-distributed plasma, which therefore modifies the transport coefficients in the previous studies of the weakly ionized plasmas with the power-law velocity distributions.

## 1 | INTRODUCTION

Collision is one of the most basic characteristics of plasma gas motion, by which the interactions along with exchanges of momentum or kinetic energy between particles are made. The average number of collisions per unit time is defined as the average collision frequency. The nature of plasma depends on the simultaneous interactions between many particles in a very unusual way and on whether Coulomb forces are directly involved in the collisions. We can distinguish two main types of collisions: one is the collisions that do not involve Coulomb forces, including collisions between two neutral particles, and collisions between neutral particles and charged particles; the other is Coulomb collisions taking place between charged particles. In the two types of collisions, there are still elastic collisions and inelastic collisions.

In plasma, "transport phenomenon" is customarily used to describe the plasma properties which are associated with the collision effects, such as diffusion, electrical conductivity, thermal conductivity, and particle traversing magnetic field etc. And many physical phenomena are described by mathematical expressions containing the collision frequencies during the transports. In addition, various processes in plasma are caused by the impact ionization and electron-ion recombination,[1] for example, the charge transfer in non-Coulomb inelastic collisions, the capture of electrons, the electromagnetic radiation and absorption in Coulomb inelastic collisions etc.[2,3] Moreover, the energy redistributions between different constitutive elements are also made by the particle-particle and particle-photon interactions in the plasma. Because many physical properties of plasma depend on collisions and collision frequency,[1-3] there have been a lot of studies of effects of the collision and collision frequency on plasma properties, such as the effect of the collision frequency on the non-magnetized or magnetized plasma waves,[4-6] the effect of the collision between dust particles on the low frequency mode and instability in dusty plasmas,[7] modeling the collision processes in the $q$-distributed plasmas,[8] transverse dielectric constant of quantum collision plasma with arbitrary collision frequency,[9] the effect of neutral collisions on instability of current-driven electrostatic ion cyclotron,[10] the effective ion-neutral collision frequency in the flowing plasmas,[11] the collision times in the plasma with super-thermal particles,[12] and the electron-ion collision frequency and the mean free path in the plasmas described by the $\kappa$-distribution etc.[13]

Almost all the investigations on the collision frequencies in plasma physics have been made so far on the basis of the traditional kinetic theory, which means that the plasma particles have to be a Maxwell-Boltzmann distribution. In fact, for the transport processes of nonequilibrium complex plasmas, the average collision frequencies depend on the velocity/energy distributions of the particles, but are often assumed to be a constant.[14-19] As we known, non-Maxwellian and/or power-law velocity and/or energy distributions are ubiquitous in many nonequilibrium complex plasmas. For example, the famous $\kappa$- and $\kappa$-like velocity/energy distributions exist widely in astrophysical and space plasmas.[20-24] In solar flares, the $\kappa$-distribution of electrons can be determined by the balance between diffusive acceleration and collisions,[25] and the $\kappa$-distribution has a significant impact on the resulting optically thin spectra arising from collision-dominated astrophysical plasmas.[26] In plasma, the $\kappa$-distribution was first observed by Vasyliunas in 1968 on the velocity distribution of electrons in the sheet of the magnetosphere.[27] But, in the past for a long time, the $\kappa$-distribution and the associated studies had not found their basis of statistical mechanics. Now it is known that the $\kappa$-distributed plasmas can be studied under the framework of nonextensive statistical mechanics.

In recent years, nonextensive statistical mechanics has attracted great attention of scientists. It has been widely used to study complex systems, such as self-gravitational astrophysics systems,[28-31] astrophysical and space plasmas,[32-34] chemical reaction systems [35,36] and biological systems etc.[37-39] In nonextensive statistics, the velocity distribution function can be expressed as the power-law $q$-distribution,[40,41]

$$f_q(\mathbf{v},\mathbf{r}) = B_q n(\mathbf{r}) \left(\frac{m}{2\pi k_B T(\mathbf{r})}\right)^{\frac{3}{2}} \left[1-(1-q)\frac{m\mathbf{v}^2}{2k_B T(\mathbf{r})}\right]^{\frac{1}{1-q}}, \qquad (1)$$

where $q$ is a nonextensive parameter whose deviation from unity represents the degree of nonextensivity, $T(\mathbf{r})$ is the temperature, $n(\mathbf{r})$ is the number density of particles, $m$ is the mass of particle, $k_B$ is the Boltzmann constant, and $B_q$ is the $q$-dependent normalized constant given by

$$B_q = \begin{cases} (1-q)^{\frac{1}{2}}(3-q)(5-3q)\dfrac{\Gamma\left(\dfrac{1}{2}+\dfrac{1}{1-q}\right)}{4\Gamma\left(\dfrac{1}{1-q}\right)}, & \text{for } 0<q\leq 1. \\[2ex] (q-1)^{\frac{3}{2}}\dfrac{\Gamma\left(\dfrac{1}{q-1}\right)}{\Gamma\left(\dfrac{1}{q-1}-\dfrac{3}{2}\right)}, & \text{for } 1\leq q < \dfrac{5}{3}. \end{cases}$$

The $q$-distribution (1) becomes a Maxwell velocity distribution only when we take the limit of $q\rightarrow 1$. For the nonequilibrium complex plasma, the $q$-parameter with a clear physical explanation was determined by the relation [41]

$$k_B \nabla T = e(1-q)\nabla \varphi_c, \qquad (2)$$

where $\varphi_c$ is a Coulombian potential and $e$ is the charge of an electron. Therefore, the $q$-distribution (1) for the $q$-parameter different from unity describes a nonequilibrium stationary-state in the complex plasma with the temperature gradient and the Coulomb interactions. For more complex plasma with a magnetic field, the $q$-parameter is found to satisfy the relation [42]

$$k_B \nabla T = (1-q)e\left(\nabla \varphi_c - c^{-1}\mathbf{u} \times \mathbf{B}\right), \qquad (3)$$

where $\mathbf{B}$ is the magnetic induction, $\mathbf{u}$ is the overall bulk velocity of the plasma and $c$ is the light speed. Here in the $q$-distribution (1), we have assumed the case $\mathbf{u}=0$.

The $q$-distribution (1) represents the type of power-law velocity distributions in nonequilibrium complex systems. Especially in complex plasma system, if we make the following parameter replacements:[43]

$$2T = (7-5q)\tilde{T} \quad \text{and} \quad (q-1)^{-1} = \kappa+1, \qquad (4)$$

the $q$-distribution (1) exactly becomes the κ-distribution observed in astrophysical and space plasmas, where $\tilde{T}$ is the temperature in the κ-distribution, which is the physical temperature of the plasma because the κ-distribution is an empirical function based on the experimental observations. The physical temperature of the $q$-distribution (1) is not $T$, but it is $\tilde{T}=2T/(7-5q)$, which was well given the physical meaning of temperature in Ref.[43]. In other words, the astrophysical and space complex plasmas with the κ-distribution can be well studied under the framework of nonextensive statistics.

Weakly ionized plasma is a kind of complex systems. Many transport coefficients of the charged particles, such as the electrical conductivity, the diffusion coefficient and the viscosity coefficient, in the weakly ionized plasmas with the $q$-distribution have been studied in nonextensive statistics.[15,16] And these coefficients are all related to the average collision frequency between electrons and neutral particles. As usual, in the previous studies of the transport coefficients of charged particles in the weakly ionized plasmas with the $q$-distributions, the collision frequency was assumed to be a constant. However, the average collision frequency between electrons and neutral particles may depend on the velocity $q$-distribution of electrons and thus depend on the $q$-parameter, which will modify the transport coefficients in the plasma. Based on this consideration, here we will study the average collision frequency between electrons and neutral particles in the weakly ionized plasmas with the $q$-distribution in nonextensive statistics.

The layout of this manuscript is as follows: Section 2 studies the average collision frequencies of electron-neutral-particle in the weakly ionized and $q$-distributed plasmas. Section 3 gives the numerical analyses of the $q$-dependent collision frequencies and the related transport coefficients. Finally, Section 4 gives the conclusion.

## 2 | THE COLLISION FREQUNCIE OF ELECTRON-NEUTRAL-PARTICLE IN THE PLASMA WITH POWER-LAW VELOCITY $q$-DISTRIBUTION

In weakly ionized plasma, when we study the electron collisions, we only consider the collisions between electrons and neutral particles. When the collision is taking place between electrons and neutral particles, they both can be regarded as a hard sphere, here we consider the hard sphere "billiard ball" collision model.[2] Because the velocities of electrons are generally much more than those of neutral particles, and the mass of an electron is much less than that of the neutral particle, the neutral particles can be considered to be relatively a static state. In this case, traditionally, the average collision frequency of electrons-neutral-particles is defined [2] as

$$\bar{v}_{en} = N_n \int v \sigma_{en}(v) f_e(\mathbf{v}) d\mathbf{v}, \qquad (5)$$

where $f_e(\mathbf{v})$ is a velocity distribution of the electrons in the plasma, $\sigma_{en}(v)$ is the collision cross-section, $N_n$ is the number density of the neutral particles. In the hard sphere "billiard ball" collision model, the collision cross-section can be a constant and it is expressed as

$$\sigma_{en} = \pi(r_n + r_e)^2, \tag{6}$$

where $r_e$ and $r_n$ are the radius of electrons and neutral particles respectively. Then, the average collision frequency (5) is rewritten as

$$\overline{v}_{en} = N_n \sigma_{en} \int v f_e(\mathbf{v}) d\mathbf{v}, \tag{7}$$

Traditionally, if the plasma is assumed to follow a Maxwellian velocity distribution, the average collision frequency of electrons-neutral particles can be expressed [2] as

$$\overline{v}_{en} = N_n \sigma_{en} \sqrt{\frac{8k_B T_e}{\pi m_e}} \approx N_n \pi r_n^2 \sqrt{\frac{8k_B T_e}{\pi m_e}}, \tag{8}$$

where the approximation $r_e \ll r_n$ has been used, $T_e$ and $m_e$ are the electron temperature and mass, respectively.

However, for some nonequilibrium complex plasmas, the plasma particles are not a Maxwellian velocity distribution, in particular when the particles have the power-law velocity $q$-distributions, the average collision frequency (8) of the electrons-neutral particles has to be modified. Now we assume that the electrons in the nonequilibrium complex plasmas satisfy the velocity $q$-distribution (1), and then we recalculate the average collision frequency.

For $1 < q < 5/3$, by using the $q$-distribution (1), the average collision frequency is calculated (see Appendix) as

$$\begin{aligned}\overline{v}_{q,en} &= N_n \sigma_{en} \int v f_q(\mathbf{v}) d\mathbf{v} = N_n \sigma_{en} B_q \left(\frac{m_e}{2\pi k_B T_e}\right)^{3/2} \int v \left[1-(1-q)\frac{m_e \mathbf{v}^2}{2k_B T_e}\right]^{1/(1-q)} d\mathbf{v} \\ &= 4\pi N_n \sigma_{en} B_q \left(\frac{m_e}{2\pi k_B T_e}\right)^{3/2} \int_0^\infty v^3 \left[1-(1-q)\frac{m_e \mathbf{v}^2}{2k_B T_e}\right]^{1/(1-q)} dv \\ &= N_n \sigma_{en} \sqrt{\frac{8k_B T_e}{\pi m_e (q-1)}} \frac{\Gamma\left(\frac{1}{q-1}-2\right)}{\Gamma\left(\frac{1}{q-1}-\frac{3}{2}\right)}, \quad 1 < q < \frac{3}{2}.\end{aligned} \tag{9}$$

For $0 < q < 1$, because there is a thermal cutoff on the maximum value allowed for the velocity in the $q$-distribution (1), we denote

$$v_{\max} = \sqrt{\frac{2k_B T_e}{m_e(1-q)}}. \tag{10}$$

Therefore, by using the $q$-distribution (1), the average collision frequency is calculated (see Appendix) as

$$\begin{aligned}\overline{v}_{q,en} &= N_n \sigma_{en} \int v f(\mathbf{v}) d\mathbf{v} \\ &= 4\pi N_n \sigma_{en} B_q \left(\frac{m_e}{2\pi k_B T_e}\right)^{3/2} \int_0^{v_{\max}} v^3 \left[1-(1-q)\frac{m_e \mathbf{v}^2}{2k_B T_e}\right]^{1/(1-q)} dv \\ &= N_n \sigma_{en} \sqrt{\frac{8k_B T_e}{(1-q)\pi m_e}} \frac{\Gamma\left(\frac{5}{2}+\frac{1}{1-q}\right)}{\Gamma\left(3+\frac{1}{1-q}\right)}, \quad 0 < q < 1.\end{aligned} \tag{11}$$

By combining Eq. (9) with Eq. (11), we can obtain the average collision frequency of electrons-neutral-particles in the plasma with the power-law velocity $q$-distribution, namely,

$$\bar{v}_{q,en} = N_n \pi r_n^2 \sqrt{\frac{8k_B T_e}{\pi m_e}} \begin{cases} \sqrt{\frac{1}{(1-q)}} \dfrac{\Gamma\left(\frac{5}{2}+\frac{1}{1-q}\right)}{\Gamma\left(3+\frac{1}{1-q}\right)}, & 0<q<1, \\ \sqrt{\frac{1}{(q-1)}} \dfrac{\Gamma\left(\frac{1}{q-1}-2\right)}{\Gamma\left(\frac{1}{q-1}-\frac{3}{2}\right)}, & 1<q<\frac{3}{2}. \end{cases} \quad (12)$$

In the above expression (12), it is clear that, the nonextensive parameter $q \neq 1$, i.e. the power-law index, has a very significant effect on the average collision frequency of electrons-neutral-particles in the complex plasma with the velocity $q$-distribution, and when we take the limit $q \to 1$, the collision frequency (12) recovers the traditional form (8) in the plasma with a Maxwell velocity distribution.

Using the average collision frequency (12) which depends on the $q$-parameter, we can modify some transport coefficients obtained in the previous studies. For example, in the weakly ionized plasma with the power-law velocity $q$-distributions, when the collision frequency between the electrons and the neutral particles was assumed to be constant, the first viscosity coefficient [15] and the diffusion coefficient [16] of the electrons were expressed by

$$\eta_{q,e} = \frac{2n_e k_B T_e}{v_e (7-5q)}, \quad 0<q<\frac{7}{5}, \quad (13)$$

$$D_{q,e} = \frac{2k_B T_e}{v_e m_e (7-5q)}, \quad 0<q<\frac{7}{5}, \quad (14)$$

where $v_e$ is the collision frequency. With the new collision frequency (12) to replace this constant collision frequency in a Maxwell distribution, the above two transport coefficients can be modified as the following forms:

$$\eta_{q,e} = \frac{n_e}{N_n r_n^2} \sqrt{\frac{k_B T_e m_e}{2\pi}} \begin{cases} \dfrac{(1-q)^{1/2}}{(7-5q)} \dfrac{\Gamma\left(3+\frac{1}{1-q}\right)}{\Gamma\left(\frac{5}{2}+\frac{1}{1-q}\right)}, & 0<q<1 \\ \dfrac{(q-1)^{1/2}}{(7-5q)} \dfrac{\Gamma\left(\frac{1}{q-1}-\frac{3}{2}\right)}{\Gamma\left(\frac{1}{q-1}-2\right)}, & 1<q<\frac{7}{5} \end{cases} \quad (15)$$

$$D_{q,e} = \frac{1}{N_n r_n^2} \sqrt{\frac{k_B T_e}{2m_e \pi}} \begin{cases} \dfrac{(1-q)^{1/2}}{(7-5q)} \dfrac{\Gamma\left(3+\frac{1}{1-q}\right)}{\Gamma\left(\frac{5}{2}+\frac{1}{1-q}\right)}, & 0<q<1 \\ \dfrac{(q-1)^{1/2}}{(7-5q)} \dfrac{\Gamma\left(\frac{1}{q-1}-\frac{3}{2}\right)}{\Gamma\left(\frac{1}{q-1}-2\right)}, & 1<q<\frac{7}{5} \end{cases} \quad (16)$$

It is clear that due to the new average collision frequency which depends on the $q$-parameter, the dependences of the transport coefficients of the charged particles in the weakly ionized plasma on the $q$-parameter are significantly different from the previous results where the collision frequency was assumed a constant. Therefore, based on the present consideration, the transport coefficients should describe better the transport processes in the weakly ionized plasma with the velocity $q$-distribution in nonextensive statistics.

In next section, we will numerically analyze the roles of the $q$-parameter in the average collision frequency as well as the transport coefficients in the plasma.

## 3 | NUMERCIAL ANALYSES OF THE NONEXTENSIVE EFFECTS

In order to show the role of the nonextensive $q$-parameter in the average collision frequency of electrons-neutral-particles in the plasma with the $q$-distribution more clearly, we make numerical analyses for the expression (12). For this purpose, from Eq.(12) and Eq.(8) we write that

$$\frac{\overline{\nu}_{q,en}}{\overline{\nu}_{en}} = \begin{cases} \sqrt{\dfrac{1}{1-q}}\,\dfrac{\Gamma\left(\dfrac{5}{2}+\dfrac{1}{1-q}\right)}{\Gamma\left(3+\dfrac{1}{1-q}\right)}, & 0<q<1, \\[2em] \sqrt{\dfrac{1}{q-1}}\,\dfrac{\Gamma\left(\dfrac{1}{q-1}-2\right)}{\Gamma\left(\dfrac{1}{q-1}-\dfrac{3}{2}\right)}, & 1<q<\dfrac{3}{2}. \end{cases} \quad (17)$$

Eq.(17) is a ratio of the average collision frequency of electrons-neutral-particles in the $q$-distributed plasma to that in the Maxwell-distributed plasma, which expresses the role of the $q$-parameter in the average collision frequency.

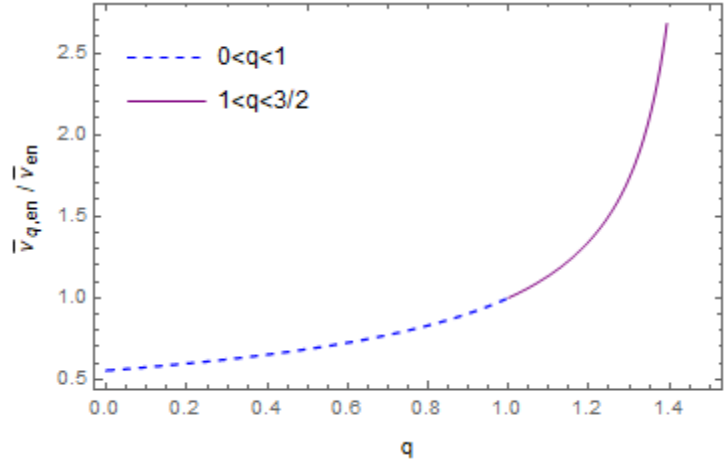

**FIGURE 1** Dependence of the average collision frequency on the $q$-parameter

In Figure 1, the numerical analysis is made on the basis on Eq.(17), which shows dependence of the average collision frequency on the $q$-parameter in the two different value ranges of the $q$-parameter: $0<q<1$ and $1<q<3/2$. It is shown that the collision frequency will increase monotonously as the $q$-parameter increases. Therefore, in the range of $0<q<1$, the average collision frequency in the $q$-distributed plasma is less than that in the Maxwell-distributed plasma, but in the range of 1< $q$ < 3/2, the average collision frequency in the $q$-distributed plasma is more than that in the Maxwell-distributed plasma.

Now we study numerically the effect of the $q$-parameter on the modified first viscosity coefficient and diffusion coefficient in Eq.(15) and Eq.(16). For convenience of analyses, we can write the ratio of Eq.(15) and Eq.(16) respectively to those in the Maxwell-distributed plasma as follows:

$$\frac{D_{q,e}}{D_{1,e}} = \frac{\eta_{q,e}}{\eta_{1,e}} = \begin{cases} \dfrac{(1-q)^{1/2}}{(7-5q)} \dfrac{\Gamma\left(3+\dfrac{1}{1-q}\right)}{\Gamma\left(\dfrac{5}{2}+\dfrac{1}{1-q}\right)}, & 0 < q < 1, \\[2ex] \dfrac{(q-1)^{1/2}}{(7-5q)} \dfrac{\Gamma\left(\dfrac{1}{q-1}-\dfrac{3}{2}\right)}{\Gamma\left(\dfrac{1}{q-1}-2\right)}, & 1 < q < \dfrac{7}{5}, \end{cases} \quad (18)$$

where $D_{1,e}$ and $\eta_{1,e}$ are respectively the diffusion coefficient and the first viscosity coefficient of the electrons in the weakly ionized and Maxwell-distributed plasma. Namely,

$$D_{1,e} = \frac{1}{N_n r_n^2} \sqrt{\frac{k_B T_e}{8\pi m_e}}, \quad (19)$$

$$\eta_{1,e} = \frac{n_e}{N_n r_n^2} \sqrt{\frac{k_B T_e m_e}{8\pi_e}}. \quad (20)$$

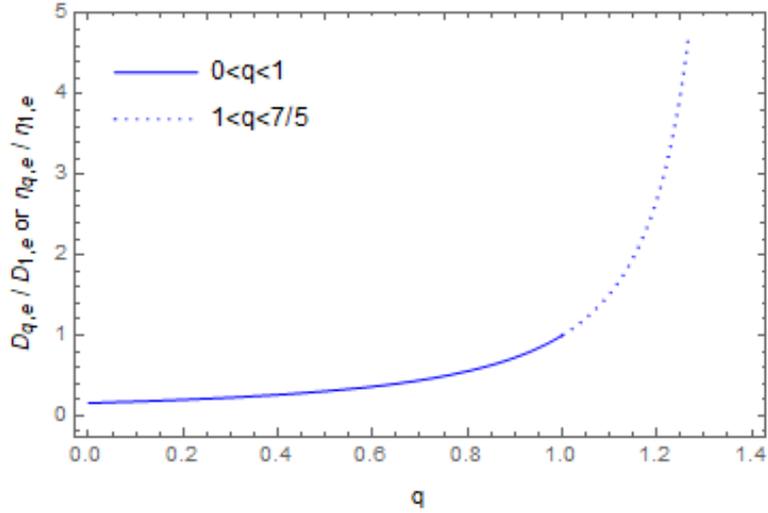

**FIGURE 2**  Dependence of the new first viscosity coefficient and diffusion coefficient on the $q$-parameter.

In Figurer 2 we give numerical results of the two transport coefficients based on (18) in two different range of values of the $q$-parameters: 0< $q$ <1 and 1< $q$ <7/5. It is shown that the new first viscosity coefficient and the diffusion coefficient will both increase monotonously as the $q$-parameter increases. In the range of 0< $q$ <1, the new first viscosity coefficient and diffusion coefficient in the $q$-distributed plasma are both less than those in the Maxwell-distributed plasma, but in the range of 1< $q$ < 7/5 they are both more than those in the Maxwell-distributed plasma.

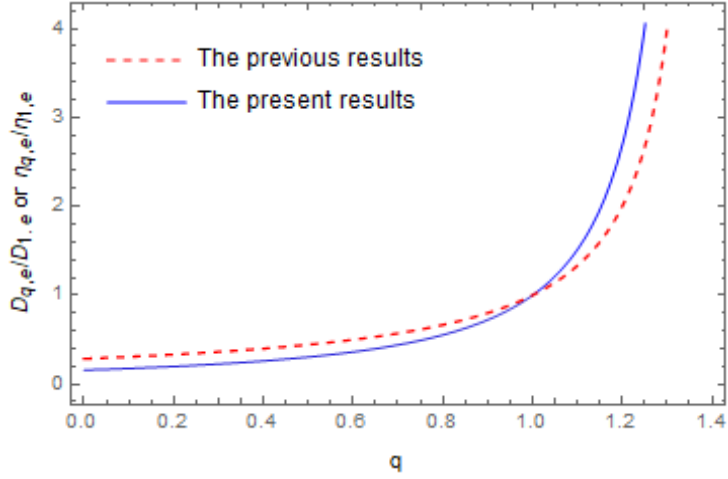

**FIGURE 3** The present transport coefficients with the $q$-dependent collision frequency compared with the previous transport coefficients with the constant collision frequency.

Further, in order to compare the present transport coefficients having the $q$-dependent collision frequency with the previous transport coefficients having the constant collision frequency more clearly, in the Figure 3 we plotted the two results both as a function of the $q$-parameter, where the solid line represents the present results based on (18) with the $q$-dependent collision frequency, and the dashed line represents the previous results with the constant collision frequency in Refs.[15,16].

It is shown that in the range of 0< $q$ <1, the two transport coefficients of electrons in the present results are both less than those in the previous results, but in the rage of 1< $q$ < 7/5, the present results are both more than those in the previous results. The average collision frequency depends strongly on the $q$-parameter and therefore has a significant effect on the transport coefficients of electrons in the weakly ionized plasma with the $q$-distributions. This effect should be taken into consideration in the study of the plasma transport coefficients.

## 4 | CONCLUSION

In conclusion, we have studied the average collision frequency of electrons- neutral-particles collision in the weakly ionized plasma with the power-law velocity $q$-distribution (1). We derived the $q$-dependent formulation of the average collision frequency of electrons-neutral-particles collisions in the plasma, which is expressed by Eq. (12). We find that the collision frequency depends strongly on the $q$-parameter of the $q$-distributed plasma, and in the limit $q \to 1$ the collision frequency perfectly returns to that in the Maxwell-distributed plasma. With the $q$-dependent collision frequency, the transport coefficients, such as the first viscosity coefficient and the diffusion coefficient of the electrons in previous studies,[15,16] are rewritten as Eq.(15) and Eq.(16), which thus modifies the previous transport coefficients in the weakly ionized plasma with the velocity $q$-distribution. It is shown that the average collision frequency depends strongly on the $q$-parameter and therefore has a significant effect on the transport coefficients of the electrons in the weakly ionized plasma with the power-law velocity $q$-distributions. Therefore, this effect should be added when we study the transport processes in the weakly ionized and *q*-distributed plasma.

Further, the numerical analyses show that the collision frequency between electrons and neutral particles increases monotonously as the *q*-parameter increases. In the range of $0 < q < 1$, the collision frequency in the *q*-distributed plasma is less than that in the Maxwell-distributed plasma, but in the range of $1 < q < 3/2$, the collision frequency in the *q*-distributed plasma is more than that in the Maxwell-distributed plasma. For the two transport coefficients, i.e., the first viscosity coefficient and the diffusion coefficient of electrons in the weakly ionized plasma with the velocity *q*-distribution, it is shown that in the range of $0 < q < 1$, the present results are both less than the previous results, but in the rage of $1 < q < 7/5$, the present results are both more than the previous results.

**ACKNOWLEDGMENTS**

This work was supported by the National Natural Science Foundation of China under Grant No. 11775156.

**APPENDIX**

In Equation (9), we have that

$$\bar{v}_{q,en} = 4\pi N_n \sigma_{en} B_q \left(\frac{m_e}{2\pi k_B T_e}\right)^{3/2} \int_0^\infty v^3 \left[1-(1-q)\frac{m_e v^2}{2k_B T_e}\right]^{1/(1-q)} dv, \tag{A.1}$$

where the integral is calculated as

$$\int_0^\infty v^3 \left[1-(1-q)\frac{m_e v^2}{2k_B T_e}\right]^{1/(1-q)} dv$$

$$= \frac{1}{(2-q)}\frac{2k_B T_e}{m_e} \int_0^\infty v \left[1-(1-q)\frac{m_e v^2}{2k_B T_e}\right]^{\frac{1}{1-q}+1} dv$$

$$= \frac{1}{2(2-q)(3-2q)}\left(\frac{2k_B T_e}{m_e}\right)^2, \quad 1 < q < \frac{3}{2}. \tag{A.2}$$

Substituting (A2) as well as $B_q$ in the *q*-distribution (1) into (A1), we derive that

$$\bar{v}_{q,en} = N_n \sigma_{en} \sqrt{\frac{8k_B T_e}{\pi m_e (q-1)}} \frac{\Gamma\left(\frac{1}{q-1}-2\right)}{\Gamma\left(\frac{1}{q-1}-\frac{3}{2}\right)}, \quad 1 < q < \frac{3}{2}. \tag{A.3}$$

This is Equation (9).

In Equation (11), we have that

$$\bar{v}_{q,en} = 4\pi N_n \sigma_{en} B_q \left(\frac{m_e}{2\pi k_B T_e}\right)^{3/2} \int_0^{v_{max}} v^3 \left[1-(1-q)\frac{m_e \mathbf{v}^2}{2k_B T_e}\right]^{1/(1-q)} dv, \tag{A.4}$$

where the integral is calculated as

$$\int_0^{v_{\max}} v^3 \left[1-(1-q)\frac{m_e v^2}{2k_B T_e}\right]^{1/(1-q)} dv$$

$$= -\frac{1}{(q-2)}\frac{2k_B T_e}{m_e}\int_0^{v_{\max}} v\left[1-(1-q)\frac{m_e v^2}{2k_B T_e}\right]^{\frac{1}{1-q}+1} dv$$

$$= \frac{1}{2(2-q)(3-2q)}\left(\frac{2k_B T_e}{m_e}\right)^2. \tag{A.5}$$

Substituting (A.5) as well as $B_q$ in the $q$-distribution (1) into (A.4), we derive that

$$\bar{v}_{q,en} = N_n \sigma_{en}\sqrt{\frac{8k_B T_e}{(1-q)\pi m_e}}\frac{\Gamma\left(\frac{5}{2}+\frac{1}{1-q}\right)}{\Gamma\left(3+\frac{1}{1-q}\right)}, \quad 0 < q < 1. \tag{A.6}$$

This is Equation (11).